\documentclass{emulateapj}

\shorttitle{Recoiling SMBH in SDSSJ0927+2943}
\shortauthors{Komossa et al.}

\begin{document}

\title{A recoiling supermassive black hole in the quasar SDSSJ092712.65+294344.0?} 

\author{S. Komossa, H. Zhou, H. Lu} 
\affil{Max-Planck-Institut f\"ur extraterrestrische Physik,
Postfach 1312, 85741 Garching, Germany}

\begin{abstract}
We present SDSSJ092712.65+294344.0 as the best candidate to date for a recoiling
supermassive black hole (SMBH).  
SDSSJ0927+2943 shows an exceptional optical emission-line spectrum with two
sets of emission lines: one set of very narrow emission lines,
and a second set of broad Balmer and broad high-ionization forbidden lines
which are blueshifted by 2650 km\,s$^{-1}$ relative to the set of narrow emission lines.  
This observation is most naturally explained if the SMBH was ejected from
the core of the galaxy, carrying with it the broad-line gas
while leaving behind the bulk of the narrow-line gas. 
We show that the observed properties of SDSSJ0927+2943 are
consistent with predictions and expectations  
from recent numerical relativity simulations which
demonstrate that  SMBHs
can receive kicks up to several thousand km\,s$^{-1}$ due to anisotropic emission
of gravitational waves during the coalescence of a binary.  
Our detection of a strong candidate for a rapidly recoiling SMBH implies
that kicks large enough to remove SMBHs completely from their host galaxies
do occur, with important implications for
models of black hole and galaxy assembly at the epoch of structure formation,
and for recoil models.

\end{abstract}


\section{Introduction}

The merging of two galaxies will produce a binary black hole
at the center of the newly formed galaxy.
If the two black holes do not stall, they will 
ultimately merge due to emission of gravitational
wave radiation. The gravitational waves carry away linear
momentum, causing the centre of mass of the coalescing BH system
to recoil in the opposite direction (Peres 1962, Bekenstein 1973).  
Early analytical calculations predicted that mergers of non-spinning black holes can 
attain kicks with velocities
of up to a few hundred km\,s$^{-1}$ (e.g., Fitchett \& Detweiler 1984,
Favata et al. 2004, Blanchet et al. 2005, Damour \& Gopakumar 2006),
recently confirmed by numerical simulations (e.g., Baker et al. 2006, Herrmann et al.
2007a, Gonz{\'a}lez et al. 2007a). These velocities are above
the escape velocity of dwarf galaxies, low-mass spirals, and high-redshift dark matter
halos. If many BHs were removed from their hosts in the early
history of the universe, this would have profound consequences for galaxy assembly and BH growth
in the early universe,
and would give rise to a population of interstellar and intergalactic BHs   
(e.g., Madau et al. 2004, Merritt et al. 2004, Madau \& Quataert 2004, 
Haiman 2004, Yoo \& Miralda-Escud{\'e} 2004, Volonteri \& Perna 2005, Volonteri \& Rees 2006, 
Libeskind et al. 2006). 

Recent numerical relativity simulations of certain configurations of
merging, {\em spinning} BHs
have produced much
higher recoil velocities, up to several thousand km\,s$^{-1}$
(Campanelli et al. 2007a,b, Gonz{\'a}lez et al. 2007b,
Tichy \& Marronetti 2007,
Herrmann et al. 2007b, Dain et al. 2008, Schnittman et al. 2008), scaling to an expected maximum
around 4000 km\,s$^{-1}$ (Campanelli et al. 20007a,b, Baker et al. 2008) 
for maximally spinning equal-mass binaries with anti-aligned spins 
in the orbital plane.
These kick velocities
exceed the escape velocities of
even massive elliptical galaxies (Fig. 2 of Merritt et al. 2004)
and  therefore  the new results
reinforce and enhance consequences studied earlier for the smaller
kicks, with   
potentially far-reaching implications
for the early phases of BH growth from early stellar-mass precursors
or later intermediate-mass 
precursors (Schnittman 2007, Volonteri 2007) and consequently for the frequency
of gravitational wave signals detectable with {\sl LISA} (Sesana 2007), 
for the scatter in the $M-\sigma$ relation (Libeskind et al. 2006),
and for the offsets and oscillations 
of recoiling BHs in galaxy cores (Gualandris \& Merritt 2008). 
 
The recoiling black holes will carry a fraction of nuclear gas and stars with
them (Merritt et al. 2004, 2006, Madau \& Quataert 2004, Loeb 2007). 
They would be detectable spatially in the form
of Seyfert or quasar activity offset from the galaxy core
(Madau \& Quataert 2004), or in the
form of broad emission lines kinematically offset from the narrow emission lines
(Bonning et al. 2007, Komossa et al. 2008).  
Because of the broad astrophysical implications, the search for
and actual identification of such recoiling black holes is of great
interest, and will place important constraints on BH growth during the epoch
of structure formation,
on predictions of maximum recoil velocity, and on
arguments suggesting 
that the BH spin configurations leading to maximal
recoil velocities should be rare in gas-rich mergers (Bogdanovi{\'c} et al. 2007).

Bonning et al. (2007) searched for recoiled SMBHs 
in the Sloan Digital Sky Survey (SDSS) database,
looking for systematic kinematic offsets between broad-line gas attached
to the recoiling BH, and narrow-line gas left behind. 
They did not find any promising candidate, and concluded that SMBH recoil
with large kick velocities is relatively rare. 

Here, we present the best candidate to date for a recoiling SMBH,
the quasar\\ SDSSJ092712.65+294344.0 (SDSSJ0927+2943 hereafter). 
Its unusual emission-line spectrum
matches key predictions from the recoiled-SMBH scenario.  
We use a cosmology with
$H_{\rm 0}$=70 km\,s$^{-1}$\,Mpc$^{-1}$, $\Omega_{\rm M}$=0.3
and $\Omega_{\rm \Lambda}$=0.7 throughout this Letter.

\section{Observations of SDSSJ0927+2943}

\subsection{Optical spectroscopy}

SDSSJ0927+2943 at redshift $z$=0.713 is a luminous quasar,
observed in the course of the SDSS 
(Adelman-McCarthy et al. 2007), and was found by us 
in a systematic search for active galactic nuclei (AGN) with high [OIII]
velocity shifts. 
The SDSS spectrum, corrected for the Galactic reddening of
E(B-V) = 0.021 mag, 
is displayed in Fig. 1. The underlying continuum 
spectral energy distribution (SED) was modeled as
a powerlaw with a best-fit slope 
of $\alpha=-0.1$ (where $f_{\nu}\propto \nu^{\alpha}$).
Each emission line was fit by a single Gaussian except the FeII
multiplets, which were modeled by templates built from I Zw 1
($\lambda>3500$ \AA, V{\'e}ron-Cetty et al. 2004; $\lambda<3500$ \AA,
Tsuzuki et al. 2006). The redshifts of the FeII lines
were tied either to MgII (the UV multiplets) or to broad H$\beta$ 
(the optical multiplets).
 
Two systems of strong emission lines can be identified in the spectrum, 
which we refer to as the ``red'' (r) and ``blue'' (b) systems. 
The red system consists of very
narrow emission lines (red NELs, r-NELs hereafter) of [OIII]5007, [OII]3727, [NeIII]3869, faint [NeV]3426
and Balmer lines, all of them almost unresolved
(FWHM,obs([OIII]) = 230 km\,s$^{-1}$; the widths of the narrow lines are
all very similar, and were therefore all fixed to the same value in order
to derive fluxes). 
The blue system shows classical broad Balmer and MgII2798 emission lines (BELs),
plus unusually broad NELs (blue NELs, b-NELs hereafter).   
All lines of the blue system are blueshifted by about 2650 km\,s$^{-1}$ relative to 
the r-NELs{\footnote{Since the r-NELs are all very narrow and are all
consistent with the same redshift, we assume henceforth that they define
the restframe of the system.}}(see Tab. 1 for redshifts; the value of 2650 km\,s$^{-1}$
is the shift between broad H$\beta$ and r-[OIII]).
The b-NELs show broad [NeV] with
a width of FWHM([NeV])=2080 km\,s$^{-1}$, and broad [NeIII] 
with FWHM([NeIII])=1020 km\,s$^{-1}$.
[OIII] and [OII] are present, too, with
widths of 460 km\,s$^{-1}$. The BELs appear in Balmer lines and in MgII
with FWHM(H$\beta$)=5740 km\,s$^{-1}$ and FWHM(MgII)=3530 km\,s$^{-1}$. 
Line ratios indicate AGN-like
excitation in both systems, r-NELs and b-NELs.
Emission-line properties are summarized in Tab. 1. 
 
In order to see whether any (faint) broad-line emission is
also accompanying the r-NELs, we have performed the following test. 
We have first subtracted the best-fit continuum, FeII multiplets,
and NELs from the observed spectrum, and then fit the H$\beta$
regime with two broad Gaussians. The redshift of the second Gaussian
was fixed to that of the r-NELs, and its width constrained
to be in the range $1000-5000$ km~s$^{-1}$. No successful 
fit could be obtained if a contribution of the second Gaussian
is enforced; its contribution
is always consistent with zero. 
We therefore conclude that only the b-NELs are 
accompanied by broad-line gas at the same redshift.  

The broadband SED of SDSSJ0927+2943 is rather blue with
SDSS magnitudes of $u=18.71\pm 0.02$, $g=18.36\pm 0.02$, $r=18.40\pm 0.02$,
$i=18.43\pm 0.02$, $z=18.31\pm 0.02$,   
and {\sl GALEX}{\footnote{http://galex.stsci.edu/GR2/}} (Martin et al. 2005) magnitudes of 
$NUV=18.57\pm 0.06$ and $FUV=19.49\pm 0.15$. 

Assuming that the standard broad-line region (BLR) scaling relations (Kaspi et al. 2005) hold, we
expect a BLR radius of $\sim$0.1 pc
and estimate a SMBH mass of SDSSJ0927+2943 of $M_{\rm BH}=6\,10^{8}$ M$_{\rm \odot}$
from the width of H$\beta$ and the luminosity at 5100\AA. 

\subsection{X-ray detection} 

We have searched the X-ray archives for observations of SDSSJ0927+2943.
The quasar is  
serendipitously 
located close to the edge of 
two ROSAT HRI images. The deeper exposure, of 19 ks duration, was performed
in April-May 1995. X-ray emission from SDSSJ0927+2943 is detected
with a countrate of 0.0037$\pm{0.0007}$ cts/s, which translates into
a soft X-ray luminosity of $L_{\rm x} = 5\,10^{44}$ erg/s (assuming
an X-ray spectrum with no intrinsic absorption, and with photon 
index $\Gamma_{\rm x}=-2.5$). The second HRI image was taken in November 1994
with a duration of 10 ks. SDSSJ0927+2943 is detected with a countrate of
0.0047$\pm{0.0012}$ cts/s, consistent with the other measurement.    
The X-ray detection demonstrates the presence of an inner accretion
disk (relevant for the discussion in Sect. 3.2). 

\begin{deluxetable}{lllcc}
\tabletypesize{\small}
\tablecaption{Emission-line properties of SDSSJ0927+2943.}
\tablewidth{0pt}
\tablehead{
\colhead{system} & {line} & {$z$} & {FWHM\tablenotemark{c}} & {line-ratio}}
\startdata
r-NEL & [OIII]5007   & 0.71279 & 170 & 10.1 \\
      & H$\beta$     & 0.71279\tablenotemark{a} & 170\tablenotemark{a} & 1.0\tablenotemark{b} \\
      & H$\gamma$    & 0.71279\tablenotemark{a} & 170\tablenotemark{a} & 0.4 \\
      & [NeIII]3869  & 0.71279\tablenotemark{a} & 170\tablenotemark{a} & 0.7 \\
      & [OII]3727    & 0.71279\tablenotemark{a} & 170\tablenotemark{a} & 2.6 \\
      & [NeV]3426    & 0.71279\tablenotemark{a} & 170\tablenotemark{a} & 0.3 \\
\tableline
b-NEL & [OIII]5007   & 0.69713 & 460 & 6.7 \\
      & H$\beta$     & 0.69713\tablenotemark{a} & ~460\tablenotemark{a} & 1.0\tablenotemark{b} \\
      & [NeIII]3869  & 0.69678 & 1020 & 1.8 \\
      & [OII]3727    & 0.69801 & ~460\tablenotemark{a} & 1.5 \\
      & [NeV]3426    & 0.69709 & 2080 & 4.0 \\
  BEL & H$\beta$     & 0.69770 & 5740 & 30.7 \\
      & H$\gamma$    & 0.69970 & 3880 & 10.7 \\
      & H$\delta$    & 0.69996 & 3260 & 3.8  \\
      & MgII 2798    & 0.69832 & 3530 & 29.5
\enddata
\tablenotetext{a}{fixed}
\tablenotetext{b}{normalized to 1.0}
\tablenotetext{c}{corrected for instrumental resolution}
\end{deluxetable}

\section{Discussion}

\subsection{The most unusual quasar pair known?}

Several pairs of quasars have been detected at projected separations of 3-10$^{\prime\prime}$.
While some of them can be explained by lensing, others very likely
represent real pairs (e.g., Kochanek et al. 1999; review by Komossa 2003).{\footnote{In this context,
it is interesting to mention HE0450-2958,
a system consisting
of an ultraluminous infrared galaxy (ULIRG) and 
a narrow-line Seyfert 1 (NLS1) galaxy. 
The host galaxy of the NLS1
was not detected in {\sl HST} imaging, 
suggesting it was a `naked quasar' (Magain et al. 2005).
Merritt et al. (2006) then revised downward the SMBH mass of
the NLS1 by a factor 10; the non-detection of the host galaxy is then
consistent with the observed upper limit. 
Merritt et al. (2006) argued that the 
emission-line properties 
of the NLS1 are incompatible
with the scenario that the NLS1 was actually ejected from the ULIRG,
whether the mechanism of ejection was gravitational
wave recoil or three-body interaction of a binary BH with a third BH
(Haehnelt et al. 2006, Hoffman \& Loeb 2006). Finally, Kim et al. (2007)
pointed out that the ultraluminous IR emission is more likely associated
with the NLS1 galaxy.  
}}  
Is SDSSJ0927+2943 actually a binary quasar ? We temporarily
refer to this hypothetical system as SDSSJ0927+2943A,B. 
The difference in velocity of the two
sets of emission lines 
exceeds the peculiar velocities observed
in galaxy clusters, and is too large for the two galaxies to form a bound
merger. Their redshift difference corresponds to $\sim$100 Mpc, if their
redshifts are cosmological. 
Consequently, we would then have a very unlikely projection effect, including not
just one, but two intrinsically extremely unusual AGN:
SDSSJ0927+2943A with exceptionally broad neon lines of [NeIII]
and [NeV]. 
And SDSSJ0927+2943B as one of the rare type-2 quasars with, 
in addition, exceptionally narrow emission lines
[their observed width of $\sim$230 km\,s$^{-1}$
is below that typically
observed in quasar narrow-line regions (FWHM $\approx 400-800$ km\,s$^{-1}$;
Zakamska et al. 2003)]. This rare
source would have to be projected by chance behind the other unusual AGN.   

We are therefore let to consider the alternative hypothesis
that we actually see only one AGN, its
SMBH and bound emission-line region having been ejected from the core with
a line-of-sight speed of 2650 km\,s$^{-1}$, while the bulk of the narrow-line
region (NLR) and other interstellar medium (ISM) was left behind and 
shines in narrow emission lines.  
This is the scenario actually predicted
in the context of the recent black hole recoil simulations
discussed in Sect. 1, and it is the observational signature
Bonning et al. (2007) searched for, and Komossa et al. (2008)
paid attention to, but did not detect. 

\begin{figure}
\plotone{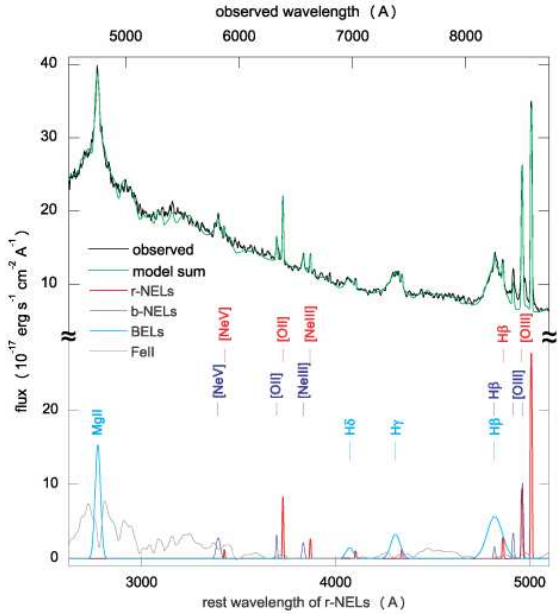}
\caption{SDSS spectrum of SDSSJ0927+2943, showing two sets of emission lines
separated by a velocity of $v \approx$ 2650 km\,s$^{-1}$. Red: red set of narrow emission lines (r-NELs),
blue and light blue: blue set of emission lines (b-NELs and BELs, respectively),
grey: FeII spectrum.
   }
\end{figure}

\subsection{A recoiling SMBH in SDSSJ0927+2943}

In this picture, SDSSJ0927+2943 underwent a merger in the past.
Upon merging, its central SMBH recoiled
taking with it the BLR
and perhaps other very high ionization gas, leaving behind the 
NLR.   
Gas with a velocity larger than the recoil velocity will
remain bound to the SMBH, and a trail of partially bound gas may form.
Accretion activity may have switched off temporarily in the course
of the binary SMBH merging, once the orbital decay time 
due to emission of gravitational waves had become smaller than
the viscous timescale of the disk 
(e.g., Liu et al. 2003, Milosavljevi{\'c} \& Phinney 2005).
After coalescence, the inner disk will re-form quickly (Loeb 2007)
and the accreting SMBH will then illuminate the bound gas, a trail of partially
bound gas and swept-up ISM, the surrounding ISM/halo 
and the left-behind bulk of the NLR gas. 

The NLR gas left behind will retain memory of the original ionization 
from the pre-merger accretion phase for only limited time,
given by the light-travel time, and the hydrogen recombination timescale, 
$t_{\rm rec} \approx 130\,T_{4}^{0.8}n_{3}^{-1}$ yr, where $T_{4}$ is
the gas temperature in 10$^4$ K, and $n_{3}$ is the gas density in units
of 10$^3$ cm$^{-3}$.
[OIII] will fade away more quickly than hydrogen lines
(Binette \& Robinson 1987).
However, once the recoiling SMBH re-forms its inner accretion disk, it will
re-illuminate parts of the  
NLR left behind (from larger distance than before), and any surrounding 
ISM.  Emission-line ratios would depend on
the average density which would be higher in the classical NLR
than in other ISM/halo gas, and on metal abundances
which will be lower in the halo.

Which aspects of this scenario do we observe in SDSSJ0927+2943 ?
The blue system of emission-lines, the BELs and b-NELs, represents gas bound
to the recoiling SMBH, seen in broad Balmer lines, 
broad MgII and broad forbidden lines, especially [NeV].
After coalescence and recoil, matter orbiting the SMBH with a 
velocity much larger than the recoil
velocity will remain bound to the SMBH (Merritt et al. 2006);
i.e. matter within a region whose
size is given by $r_{\rm b} < G M_{\rm BH} v^{-2}$. In our
case, $r_{\rm b} < 10^{18}$\,cm, which is about a factor
of 3 larger than the BLR of SDSSJ0927+2943. 
The width of [NeIII] (of the b-NEL system) of 1020 km\,s$^{-1}$ is too narrow 
for this gas to be originally bound to the SMBH
(except in case of projection effects), but
as the disk keeps accreting onto the recoiling SMBH, 
it will spread radially, and will also drive some outflows.   
The semi-broad [OII] and [OIII]
most likely have this same origin, and/or are from swept-up ISM 
in front of the path of the  
recoiling SMBH{\footnote{Based on photoionization
models (e.g., Komossa \& Schulz 1997) we expect little [OII] emission from
the gas bound to the recoiling SMBH; even though shielding geometries
might be constructed in which this actually becomes possible.}}.

The red system of emission lines, the r-NELs,
represents the NLR gas left behind, and/or ISM surrounding
the recoiling SMBH, not bound to it, but illuminated by its accretion disk. 
All r-NELs have the same profiles and  
are very narrow, and the 
emission-line ratios imply an AGN ionizing continuum. 
Can we distinguish between NLR and other ISM/halo gas,
in terms of line ratios and line profiles ? 
Line {\em ratios} would depend on density and distance.
As the recoiling SMBH moves away from the galaxy core,
illumination of the NLR from an, on average larger,
distance would decrease the degree of ionization,
while illumination of low-density ISM surrounding the recoiling SMBH would
increase the degree of ionization of the line-emitting gas. It is possible that 
we see a mix of these two processes. We do not expect to see pure
very low-density halo gas. Its high ionization parameter would lead to a
much higher degree of ionization than we actually observe, as we have
also verified by test calculations with Ferland's photoionization code {\em Cloudy}
(S. Komossa et al., in preparation).  
The density-sensitive [SII] doublet 
is outside the observed wave band,
but will be accessible with NIR spectroscopy. 

Independent evidence for the origin of
the r-NELs comes from their {\em profiles} which are very narrow. Their
width is below the stellar velocity dispersion of $\sigma_*$=260 km\,s$^{-1}$
predicted from the $M_{\rm BH}-\sigma_*$ relation (Ferrarese \& Ford 2005). 
This indicates that
the r-NEL gas illuminated by the recoiling SMBH either does
not feel the full bulge potential, 
or the local velocity field of
the illuminated gas does not reflect the full velocity dispersion.
This latter situation is expected if the recoiling SMBH
was in a disk where the velocity
pattern is predominantly rotational, so that we only
see a fraction of the full pattern.

The whole parameter space of BH recoil velocities is still
being explored (e.g., Baker et al. 2006, 2007, 
Campanelli et al. 2007a,b, Choi et al. 2007,
Gonz{\'a}lez et al. 2007a,b, Herrmann et al. 2007a,b, Pollney et al.
2007, Schnittman et al. 2008, and references therein; see Pretorius 2007 for a review). 
Several recent calculations focused on (almost) equal mass BHs
with specific spin
configurations such that kick velocities are maximized.
The line-of-sight velocity we measure is comparable to the
recoil velocities predicted by  
the runs of Gonz{\'a}les et al. (2007b; $v_{\rm kick} \approx 2500-2650$ km\,s$^{-1}$)
and by Dain et al. (2008; 3300 km\,s$^{-1}$). 
The scaling formulae of Campanelli et al. (2007a) and of Baker et al. (2008)
predict maximal recoil velocities of $\sim$3800 km\,s$^{-1}$. 
In order to reach a high kick velocity, the pre-merger BHs 
of SDSSJ0927+2943 must
have been rapidly spinning and 
of nearly equal mass; i.e. the galaxy hosting
SDSSJ0927+2943 must have undergone a major merger.    

Assuming that the recoiling SMBH carried with it an amount of
mass which is not larger than its own mass, and that 
most of that gas is ultimately available for accretion, 
we can estimate an upper limit on the duration
of the quasar activity, $t_{\rm q}$. We further assume that the gas continues
to accrete at its current rate of $L/L_{\rm Edd} \simeq 0.1$, where 
$L_{\rm Edd}$ is the Eddington luminosity, and we use a
radiative efficiency of $\eta$=0.37
which is appropriate for rapidly
spinning BHs. This implies an upper limit on the lifetime
of quasar activity of $t_{\rm q} \approx 10^{9}$ yr.  
If the space velocity of the recoiling SMBH was close to the 
maximum possible velocity, it would reach a projected 
separation from the core on the order of a few kiloparsec within $\sim$10$^{6}$ yr.      

Future {\sl HST} imaging of SDSSJ0927+2943 will be valuable to distinguish whether the 
host galaxy shows
a distorted morphology as would be expected if the merger was recent. 
With its spatial resolution of 0.1$^{\prime\prime}$, {\sl HST} would allow
resolution of a spatial scale of $\sim$1 kpc in the galaxy
and measurement of a corresponding  offset of the accreting SMBH from the galaxy core. 
The X-ray detection of SDSSJ0927+2943 holds promise for a deep {\sl Chandra} study
of SMBH and host galaxy. 
If the recoiling SMBH is a radio emitter, high-resolution radio observations
would provide even more precise measurement of its location. 

In summary, SDSSJ0927+2943 is the best candidate to date
for a recoiling supermassive black hole. Its spectrum
shows the characteristic signature of two separate
emission-line systems, kinematically offset by 
$\sim$2650 km\,s$^{-1}$ -- broad lines from gas bound
to the recoiling hole, and narrow lines from gas left behind.  
Further study of this source
and detection of similar ones will provide important information on BH recoil,
relevant timescales, and the frequency of BHs removed
from their host galaxies, and therefore on simulations
of galaxy formation and evolution,
and on the question how many BHs grew early by merging.    

\acknowledgments
We thank D. Merritt and G. Hasinger for illuminating discussions.
We have made use of the SDSS, {\sl GALEX} and {\sl ROSAT} database.


\begin{thebibliography}{}

\bibitem[Adelman-McCarthy et al. 2007]{SDSS-DR} Adelman-McCarthy, J.K., et al. 2007, ApJS, 172, 634    

\bibitem[Baker et al. 2006]{bakeretal06} Baker, J.G., et al. 2006, \apj, 653, L93 

\bibitem[Baker et al. 2007]{baker07} Baker, J.G., et al. 2007, \apj, 668, 1140 

\bibitem[Baker et al. 2008]{baker08} Baker, J.G., et al. 2008, arXiv:0802.0416v1

\bibitem[Bekenstein 1973]{bekenstein73} Bekenstein, J.D. 1973, \apj, 183, 657  

\bibitem[Binette and Robinson 1987]{binette87} Binette, L., \& Robinson, A. 1987, \aap, 177, 11

\bibitem[Blanchet et al. 2005]{blanchet05} Blanchet, L., et al. 2005, \apj, 635, 508 

\bibitem[Bogdanovic et al. 2004]{bogdanovic04} Bogdanovi{\'c}, T., et al. 2007, \apj, 661, L147  

\bibitem[Bonning et al. 2007]{bonning07} Bonning, E.W., et al. 2007, \apj, 666, L13 

\bibitem[Campanelli et al. 2007a]{campanelli07a} Campanelli, M., et al. 2007a, \apj, 659, L5  

\bibitem[Campanelli et al. 2007b]{campanelli07b} Campanelli, M., et al. 2007b, Phys. Rev. Lett., 98, 1102 

\bibitem[Choi et al. 2007]{choi07} Choi, D.-I., et al. 2007, Phys. Rev. D., 76, 104026

\bibitem[Dain et al. 2008]{dain08} Dain, S., Lousto, C., Zlochower, Y., 2008, arXiv:0803.0351v1 

\bibitem[Damour and Gopakumar 2006]{damour06} Damour, T., Gopakumar, A. 2006, Phys. Rev. D., 73, 124006

\bibitem[Favata et al. 2004]{favata04} Favata, M., et al. 2004, \apj, 607, L5 

\bibitem[Ferrarese and Ford 2005]{ferrarese05} Ferrarese, L., \& Ford, H. 2005, Space Science Reviews, 116, 523 

\bibitem[Fitchett and Detweiler 1984]{fitchetdetweiler84} Fitchett, M.J., \& Detweiler, S. 1984, \mnras, 211, 933 

\bibitem[Gonzales et al. 2007a]{gonz07a} Gonz{\'a}les, J.A., et al. 2007a, Phys. Rev. Lett., 98, 091101   

\bibitem[Gonzales et al. 2007b]{gonzalesetal07b} Gonz{\'a}les, J.A., et al. 2007b, Phys. Rev. Lett., 98, 231101  

\bibitem[Gualandris et al. 2008]{gua08} Gualandris, A., \& Merritt, D. 2008, \apj, in press 

\bibitem[Haiman 2004]{haiman04} Haiman, Z. 2004, \apj, 613, 36 

\bibitem[Haehnelt et al. 2005]{haehneltetal05} Haehnelt, M., Davies, M.B., \& Rees, M.J. 2006, \mnras, 366, L22

\bibitem[Herrmann et al. 2007a]{herr07a} Herrmann, F., et al. 2007a, Class. Quantum Grav., 24, S33 

\bibitem[Herrmann et al. 2007b]{herrmann07b} Herrmann, F., et al. 2007b, Phys. Rev. D, 76, 084032 

\bibitem[Hoffman and Loeb 2006]{hoffmanandloeb06} Hoffman, L., \& Loeb, A. 2006, \apj, 638, L75 

\bibitem[Kaspi et al. 2005]{kaspi05} Kaspi, S., et al.
            2005, \apj, 629, 61 

\bibitem[Kim et al. 2007]{kim07} Kim, M., et al. 2007, \apj, 658, 107 

\bibitem[Kochanek et al. 1999]{kochanek99} Kochanek, C.S., et al. 1999, \apj, 510, 590  

\bibitem[Komossa 2003]{komossa03} Komossa, S. 2003, AIP Conf. Proc., 686, 161  

\bibitem[Komossa and Schulz 1997]{koschu97} Komossa, S., \& Schulz, H. 1997, \aap, 323, 31

\bibitem[Komossa et al. 2008]{komossa08} Komossa, S., et al. 2008, ApJ, in press (arXiv:0803.0240) 

\bibitem[Libeskind et al. 2006]{libeskind06} Libeskind, N.I., et al. 2006, \mnras, 368, 1381 

\bibitem[Liu 2004]{liu04} Liu, F. et al., 2003, \mnras, 340, 411 

\bibitem[Loeb 2007]{loeb07} Loeb, A. 2007, Phys. Rev. Lett., 99, 041103  

\bibitem[Madau et al. 2004]{madau04} Madau, P., et al. 2004, \apj, 604, 484  

\bibitem[Madau and Quataert 2004]{madauquataert04} Madau, P., \& Quataert, E. 2004, \apj, 606, L17

\bibitem[Magain et al. 2005]{magain05} Magain, P., et al. 2005, Nature, 437, 381 

\bibitem[Martin et al. 2005]{martin05} Martin, D.C., et al. 2005, \apj, 619, L1

\bibitem[Merritt et al. 2004]{merritt04} Merritt, D., et al. 2004, \apj, 607, L9 

\bibitem[Merritt et al. 2006]{merritt06} Merritt, D., et al. 2006, \mnras, 367, 1746 

\bibitem[Milosavljevic \& Phinney 2005]{milos05} Milosavljevi{\'c}, M., \& Phinney, S. 2005, \apj, 622, L93 

\bibitem[Peres 1962]{peres62} Peres, A. 1962, Phys. Rev., 128, 2471     

\bibitem[Pollney et al. 2007]{pollney} Pollney, D., et al. 2007, Phys. Rev. D, 76, 124002 

\bibitem[Pretorius 2007]{pretorius07} Pretorius, F. 2007, arXiv:07101338 

\bibitem[Schnittman 2007]{schnitt07} Schnittman, J.D. 2007, \apj, 667, L133  

\bibitem[Schnittman et al. 2008]{schnittmanetal07} Schnittman, J.D., et al. 2008, Phys. Rev. D, in press 

\bibitem[Sesana 2007]{sesana07} Sesana, A. 2007, \mnras, 382, L6 

\bibitem[Tichy and Marronetti]{tichy07} Tichy, W., \& Marronetti, P. 2007, Phys. Rev. D, 76, 061502 

\bibitem[Tsuzuki et al. 2006]{tsuzuki06} Tsuzuki, Y., et al. 2006, \apj, 650, 57  

\bibitem[Veron-Cetty et al. 2004]{veron04} V{\'e}ron-Cetty, M.-P., et al. 2004, \aap, 417, 515

\bibitem[Volonteri and Perma 2005]{volonteriperma2005} Volonteri, M., \& Perna, R. 2005, \mnras, 358, 913 

\bibitem[Volonteri and Rees 2006]{volonterirees} Volonteri, M., \& Rees, M.J. 2006, \apj, 650, 669 

\bibitem[Volonteri 2007]{volonteri07} Volonteri, M. 2007, \apj, 663, L5  

\bibitem[Yoo and Miralda-Escude 2004]{yoo04} Yoo, J., \& Miralda-Escud{\'e}, J. 2004, \apj, 614, L25 

\bibitem[Zakamska et al. 2003]{zakamska03} Zakamska, N.L., et al. 2003, \aj, 126, 2125 

\end{thebibliography}
\end{document}